# Four atomic optical energy levels as a two qubit quantum computer register

Vladimir L. Ermakov, Alexander R. Kessel and Vitaly V. Samartsev

Physico-Technical Institute, Russian Academy of Sciences, Kazan 420029, Russia

*e-mail:* ermakov@sci.kcn.ru

## ABSTRACT

It is proposed to use four atomic optical energy levels as a two qubit quantum register. A single $Pr^{3+}$ atom in a monocrystal $LaF_3$ subjected to resonant laser irradiation is used as an example to illustrate the implementation of the universal set of quantum gates. The equilibrium state of this physical system is a desirable input state for quantum computation and therefore there is no need for its special preparation procedure.

## 1. INTRODUCTION

The physical system consisting of two spins 1/2 coupled by exchange interactions is used in the standard NMR implementation of two qubit quantum gates [1,2]. Such a gate operation requires an experimental possibility to turn on the exchange interaction for a calibrated time interval and, which is more difficult, to turn it off when this gate should not function. These time intervals are defined on the exchange interaction magnitude and can exceed the coherence time. Moreover to turn the exchange interactions off an application of complicated pulse sequences is necessary. It can make the duration of experiment too long.

In [3] it was shown that the usage of spins greater or equal to 3/2 for the quantum gates and quantum register implementation gives additional facilities. The main advantage is that a two qubit gate can be realized using a single particle and therefore there is no necessity in interaction between the space separated particles. Instead of this the necessary quantum gates can be realized using short RF pulses which duration is defined by RF field amplitude and is under full control of an experimenter [3].

The virtual spin formalism, developed for an NMR quantum computer implementation [3], is adapted in the present paper to a discrete optical energy level system of a single atom. An optical quantum computer model possesses at least two advantages. Firstly, the optical system equilibrium state coincides with the initial state for a standard quantum algorithm. And secondly, there is a possibility in principle to realize the gate using a single optical particle because the modern technique allows to register a separate photon.

## 2. PHYSICAL SYSTEM

To suggest a new physical system as a quantum information processing medium it is necessary:

1. To give the description of a physical object, to find a subset of suitable stationary states among the system levels and to assign one or several qubits to them.
2. To determine which physical interactions can cause transitions between this subset levels for realizing universal logic gates - one qubit rotation and two qubit controlled not.
3. To find a way to set each qubit into initial state |0> and to read out its final state after calculations.

Let us consider step by step how to realize all these requirements using a quantum system which has several discrete energy levels.

1. Let us to choose four levels $E_3>E_2>E_1>E_0$ among several optical discrete energy levels, where $E_0$ is the basic bottom energy level. Fig.1 shows the main level $^3H_4$ and several excited energy levels $^3H_6$, $^3P_0$ and $^3P_1$ of a single $Pr^{3+}$ atom in the monocrystal $LaF_3$ [4]. In general all optical terms are split by electric quadrupole interaction. Fig.1 shows the splitting of only two terms $^3H_4$ and $^3P_0$, which are necessary to implement gates in the presented formalism of this work. These terms sublevels will be denoted as $^3H_4(m_I)$, $^3P_0(m_I)$, where $m_I$ - the nuclear spin component for quadrupole sublevel. Any four depicted energy levels can be chosen as a basis for the gates implementation depending only on convenience of the experimental realization.

2. In paper [3] a formalism is developed which shows how four energy levels of a single quantum particle can comprise two quantum bits in such a way that a universal set of quantum logic gates can be constructed. The main idea is that a four dimensional space $\Gamma_I$ of these energy levels formally can be represented as a direct product $\Gamma_R \otimes \Gamma_S$ of Hilbert spaces of two virtual spins 1/2 - R and S. In the information notation the direct product basis is $|q_r,q_s>$, where qubit states q=0 and 1 correspond to virtual spin state +1/2 and -1/2. The simultaneous resonant irradiation for transitions $E_0 \leftrightarrow E_1$ and $E_2 \leftrightarrow E_3$ at the condition that rotation angles of both transitions are equal to the same value $\varphi$ executes the virtual spin S rotation through the angle $\varphi$ keeping the spin R state invariant. In the four dimensional $\Gamma_R \otimes \Gamma_S$ state space this operation is equivalent to the transformation $exp\{ - i \varphi \mathbf{1}_R \otimes \mathbf{S}_y\}$, where $\mathbf{1}_R$ is the space $\Gamma_R$ unit operator. The appropriate transitions are depicted in Fig.2a.

The similar irradiation with transitions $E_0 \leftrightarrow E_2$ and $E_1 \leftrightarrow E_3$ fulfills the virtual spin R rotation through the angle $\varphi$ leaving the spin S invariant. In the four dimensional $\Gamma_R \otimes \Gamma_S$ state space this operation is equivalent to the transformation $exp\{ - i \varphi \mathbf{R}_y \otimes \mathbf{1}_S\}$ where $\mathbf{1}_S$ is the space $\Gamma_S$ unit operator (Fig.2b).

Two qubit controlled not gate - CNOT - can be realized by resonant excitation the single frequency transition $E_2 \leftrightarrow E_3$ with the rotation angle $\pi$ (Fig.2c). In this gate spin R state controls the spin S state. Using the virtual spin formalism this fact can be written as transformation -

$i [(1/2)\mathbf{1}_R + \mathbf{R}_z] \otimes (2\mathbf{S}_y) + [(1/2)\mathbf{1}_R - \mathbf{R}_z] \otimes \mathbf{1}_S$

which up to the phase factor $-i$ gives the CNOT truth table [5].

The similar excitation of the $E_1 \leftrightarrow E_3$ transition corresponds to CNOT gate in which spin S state controls the spin R state and can be written as the transformation (Fig.2d):

$\mathbf{1}_R \otimes [(1/2)\mathbf{1}_S - \mathbf{S}_z] - i(2\mathbf{R}_y) \otimes [(1/2)\mathbf{1}_S + \mathbf{S}_z]$.

The working four levels and the corresponding quantum gates realization can be chosen in several ways among the depicted in Fig.1 ion $Pr^{3+}$ energy levels. Two such possible schemes are shown in Fig.3 and Fig.4 below. The Fig.3 scheme uses the levels $^3H_4(0)$, $^3H_6(240\ nm)$, $^1D_2(592.5\ nm)$, $^3P_0(477.7\ nm)$ as the working ones. To realize all four quantum gates the irradiation of the following transitions are necessary: $^3H_4 \leftrightarrow ^3H_6$, $^1D_2 \leftrightarrow ^3P_0$, $^3H_6 \leftrightarrow ^3P_0$, $^3H_4 \leftrightarrow ^1D_2$. For this purpose the tunable laser on dye-stuffs «Cumarine-M80» and «Rodamine-6G», and also superradiation on the wave length 650 *nm* can be used.

The Fig.4 scheme uses the levels $^3H_4(0)$, $^3P_0(5/2)$, $^3P_0(3/2)$, $^3P_0(1/2)$ as the working ones. To realize all four quantum gates the irradiation of the following transitions $^3H_4 \leftrightarrow ^3P_0(1/2)$, $^3H_4 \leftrightarrow ^3P_0(3/2)$, $^3P_0(5/2) \leftrightarrow ^3P_0(3/2)$, $^3P_0(5/2) \leftrightarrow ^3P_0(1/2)$ are necessary. Let us point out that transitions between three upper levels are of radio frequency range and can be easily excited using electromagnetic or acoustic pulses. The transition $^3H_4 \leftrightarrow ^3P_0$ can be excited by the laser on dye-stuff «Cumarine-M80», and in this case the excitation of different quadrupole levels of term $^3P_0$ can be achieved by changing the laser beam polarization. In principle, the three quadrupole sublevels of the term $^3H_4$ and anyone upper term can be used for the aim of this work. However in this case the advantage of initial state preparing which is inherent to the optical levels will be lost, since all $^3H_4$ term sublevels will be occupied in the equilibrium.

3. The virtual spin formalism [3] presumes that only one among four states of a considered subsystem must be occupied to be a quantum algorithm input state. For a case of optical system it is naturally to choose the ground state as an input state since it is the only occupied level at the equilibrium. Such choice removes the initial state preparation procedure which is a big problem for other implementations.

The readout procedure will be considered for a particular case when the final state has only one occupied energy level. In this case the procedure can benefit the fact that the term $^3P_1$ of the $Pr^{3+}$ ion relaxes very fast to the ground state. Using 180 degrees resonant pulses the coherence of any three working levels can be transferred to $^3P_1$. Then, measuring the frequency of emitted light it is possible to determine which working level was occupied after calculations.

# REFERENCES

1. N. A. Gershenfeld and I. L. Chuang, "Bulk Spin Resonance Quantum Computation", *Science* **275,** pp. 350-356, 1997.

2. D. G. Cory, M. D. Price and T. F. Havel, "Nuclear magnetic resonance spectroscopy: An experimentally accessible paradigm for quantum computing", *Physica D* **120**, p.82, 1998.

3. A. R. Kessel and V. L. Ermakov, «Multiqubit spin», *JETP Letters* **70**, pp. 61-65, 1999 (quant-ph/9912047).

4. M.J. Weber, «Spontaneous emission probabilities and quantum efficiencies for excited states of $Pr^{3+}$ in $LaF_3$», *J. Chem. Phys.* **48**, No.10, pp.4774-4780, 1968.

5. C. P. Williams and S. H. Clearwater, *Explorations in Quantum Computing*, Springer, Telos, New York, 1998.

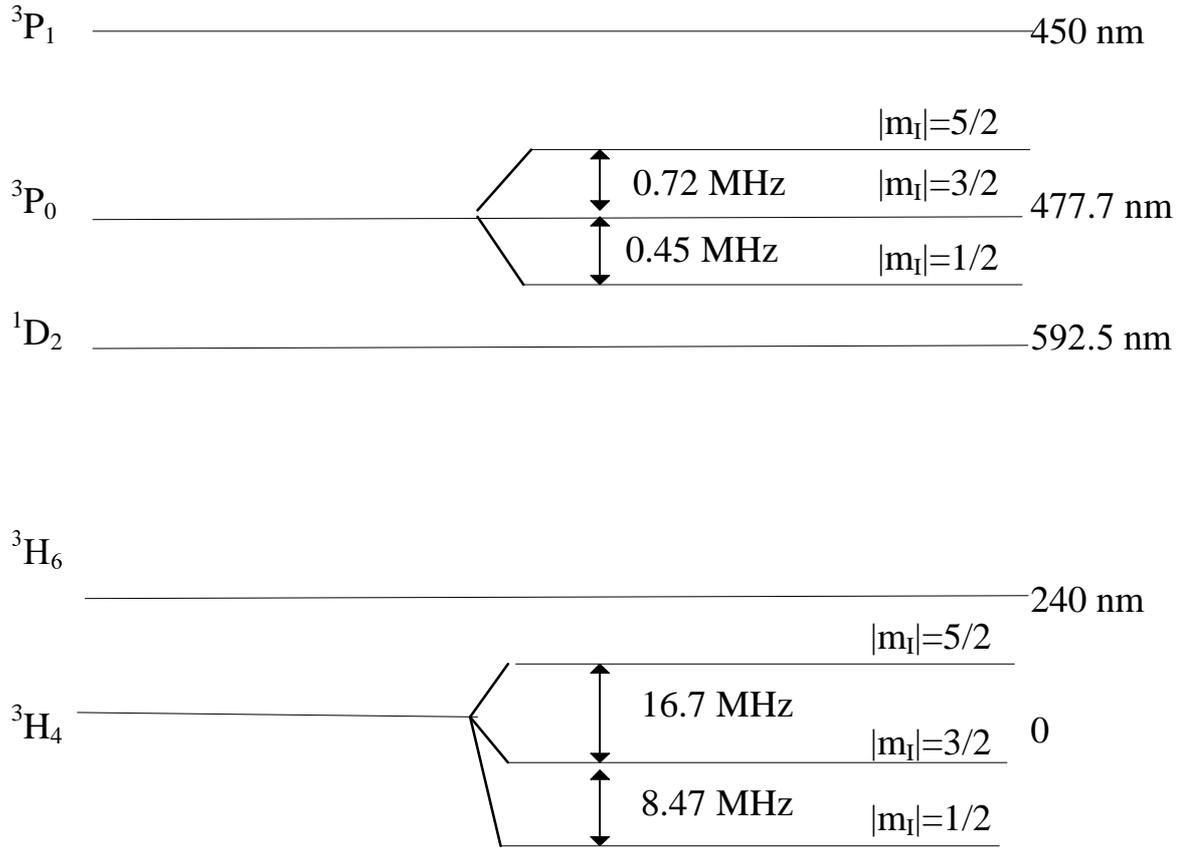

Fig.1. The ion $Pr^{3+}$ energy levels in a monocrystal $LaF_3$. The terms $^3P_0$ and $^3H_4$ quadrupole splittings are shown, which correspond to the different components $m_I$ of nucleus spin.

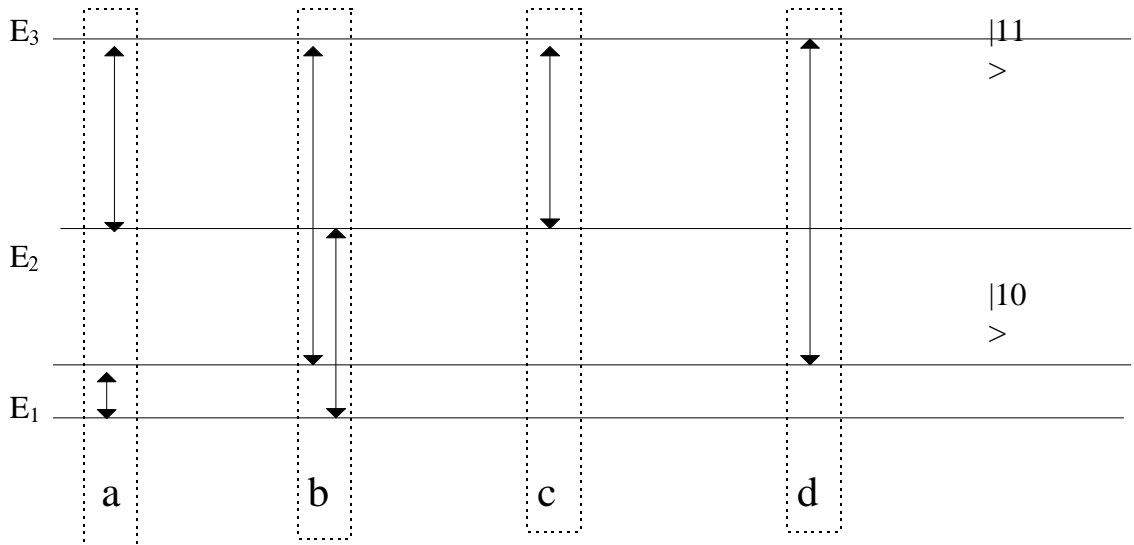

Fig.2. The excitation scheme which implements the following logic gates: a) one-qubit virtual spin R rotation; b) one-qubit virtual spin S rotation; c) CNOT- the spin R state controls the spin S state; d) CNOT- the spin S state controls the spin R state; the corresponding information notation is shown at the right side.

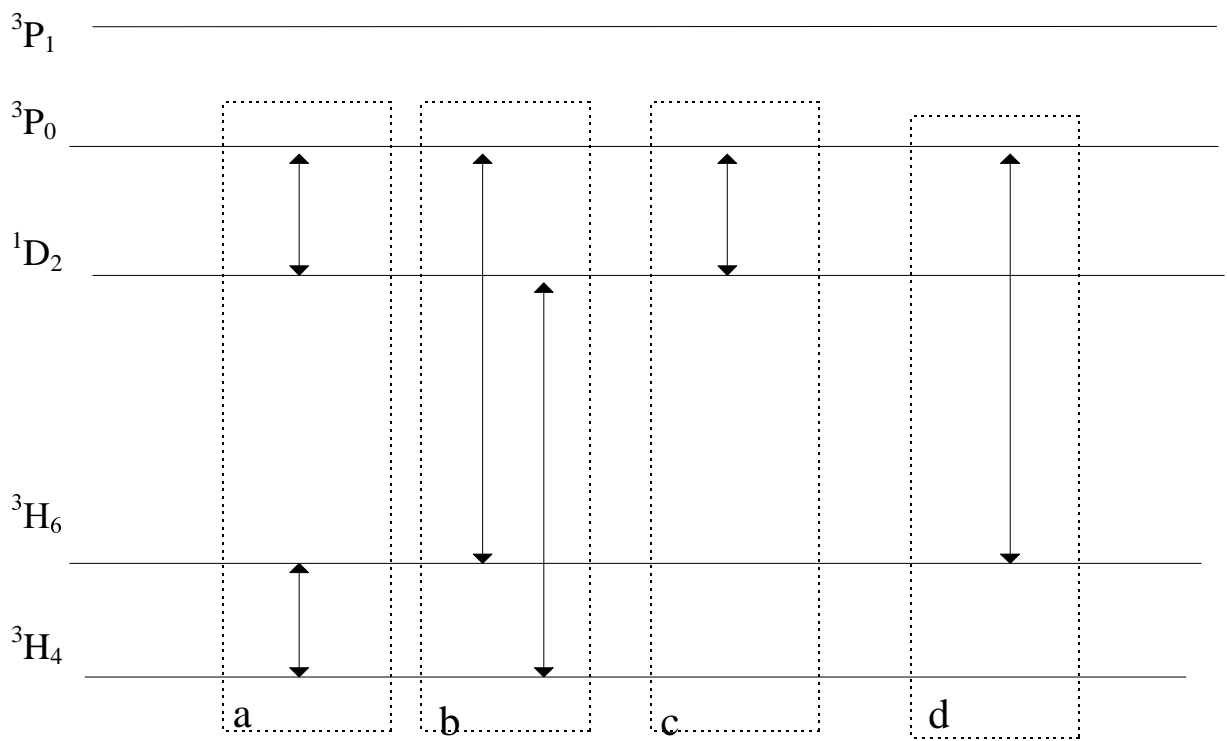

Fig.3. The excitation scheme-1 for the quantum gates implementation.

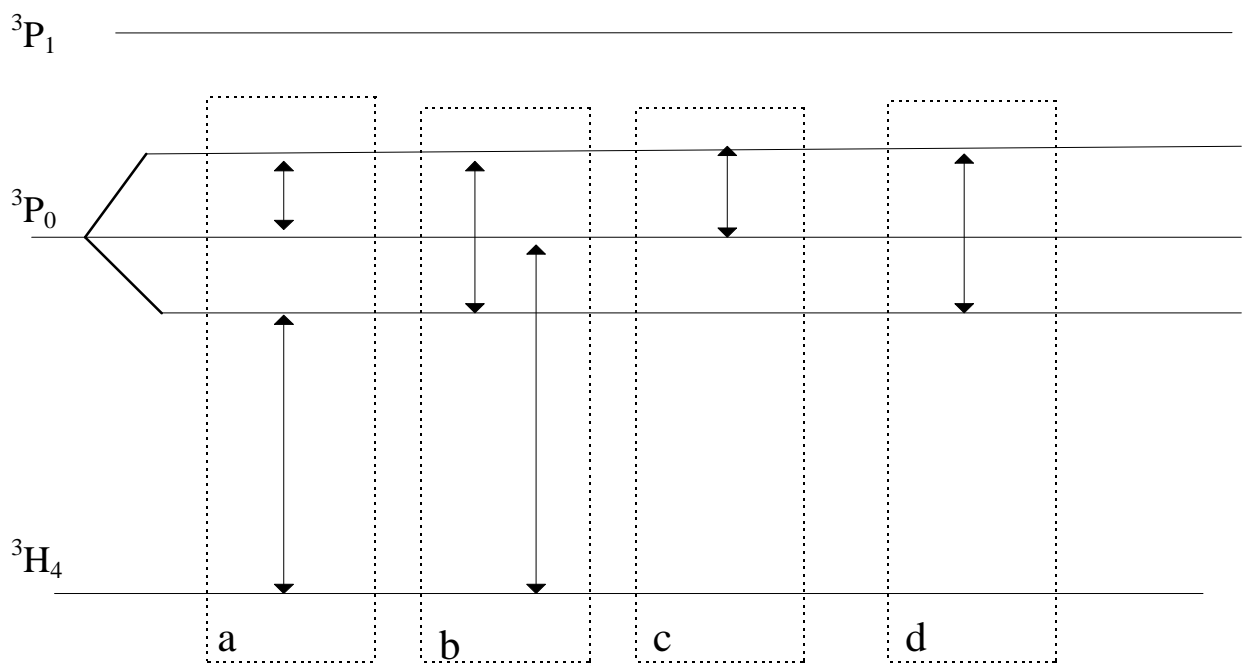

Fig.4. The excitation scheme-2 for the quantum gates implementation